# Impedance spectroscopy study on post-annealing-tuned polycrystalline $CaCu_3Ti_4O_{12}$ films: Evidence of Barrier Layer Capacitor Effects

Liang Fang, Mingrong Shen, and Zhenya Li

Department of Physics, Suzhou University, Suzhou 215006, People's Republic of China

Wenwu Cao

Materials Research Institute, The Pennsylvania State University, University Park, Pennsylvania 16802, USA

**ABSTRACT:**

In this paper, impedance spectroscopy study was performed to establish the electrical property and microstructure relations of the as-deposited and post-annealed polycrystalline CCTO films prepared on $Pt/Ti/SiO_2/Si$ (100) substrates by pulsed-laser deposition (PLD). Our results demonstrated that the as-deposited polycrystalline CCTO film was made of insulating grain boundaries with semiconducting grains, indicating that the high-dielectric-constant is attributed to the barrier layer capacitor (BLC) effects. The simple resistor-capacitor (RC) equivalent circuit and the modified constant phase element (CPE) circuit were used to describe the impedance spectroscopy, and excellent agreement between the calculated and measured curves was obtained in the CPE circuit. The resistance and capacitance of the grains and grain boundaries can be tuned by changing the annealing atmosphere and temperature. Under oxygen-absent annealing atmosphere, the electric resistances of the grain boundaries changed greatly but the resistance of the grains has almost no change. While under oxygen annealing atmosphere, the reverse happened. On the basis of this result, it is demonstrated that the origin of the semiconductivity of the grains in CCTO polycrystalline films arises from their oxygen-loss, while the grain boundaries are close to oxygen- stoicheometry.

## I. INTRODUCTION

In the past several years, there has been a revival of interest in studying high dielectric constant materials because of their potential applications in a variety of devices, such as static and dynamic random access memories. Recently, a body-centered cubic perovskite-related $CaCu_3Ti_4O_{12}$ (CCTO) compound was reported to posses an extraordinarily high static dielectric constant at room temperature with $\varepsilon_r \approx 10^5$,[1-9] which is practically frequency independent between DC and $10^6$ Hz for the temperature range between 100K and 600K. However, the dielectric constant displays a 1000-fold reduction below 100 K without any detectable change of structural phase transition, as demonstrated by neutron powder diffraction[2] and high resolution X-ray diffraction.[3] Moreover, as demanded by miniaturization of microelectronic devices, thin-film form of this materials has been prepared by several groups,[10-14] which promote not only high expectation of device applications but also challenge for better understanding of their rich physical phenomena.

Although there are many reports on the CCTO, the origin of its high dielectric constant is still not well understood. Several intrinsic physical interpretations have been given, examples include a relaxor-like slowing down of dipolar fluctuations in nanosize domains,[1] and highly polarizable relaxation modes.[3] However, without the change of crystal structure, the above mentioned effects have been questioned and some extrinsic effects have been put forth,[15] such as point or line defects[16] or Schottky barriers at the interface between sample and the electrodes.[8] M. A. Subramanian *et al.*[2,9] demonstrated that the single crystal of CCTO was highly twinned, and the twin boundaries might then be acting in a manner to create a barrier layer capacitance, thereby enhancing the dielectric constants. Such explanations were also proposed in the epitaxial CCTO thin films on oxide substrates.[9-11] D. C. Sinclair *et al.*[4,17] pointed out that, with the help of impedance spectroscopy, the insulating grain boundaries on CCTO ceramics separated semiconducting grains and they suggested that the high dielectric constant phenomenon was an extrinsic effect that was due to the barrier layer capacitors (BLC) effects.[18] However, as pointed by the authors, the origin of the semiconductivity in the grains is still to be established. In addition, there have been few reports on the detailed mechanism behind the characteristics of high dielectric constant in polycrystalline CCTO films.

In a recent work, we reported the successful deposition of CCTO thin films with high dielectric constant on Pt/Ti/SiO$_2$/Si substrates by pulsed laser deposition (PLD),[12] and it was found that the dielectric properties of the CCTO films were very sensitive to the post-annealing atmosphere and temperature.[19] However, it is not clear why the post-annealing can result in such a dramatically change of the dielectric properties. In this study, impedance spectroscopy (IS), which is a flexible experimental technique to obtain information about the electrical microstructure of materials,[20,21] was applied to measure the electrical parameters of the as-deposited and post-annealed polycrystalline CCTO films. The simple resistor-capacitor (RC) equivalent circuit and the modified constant phase element (CPE) circuit were used to interpret the experiment results. Our results provide direct evidence leading to the conclusion that the origin of the high dielectric constant in the polycrystalline CCTO thin films is due to the BLC mechanism. More importantly, the origin of the semiconductivity in the grains is demonstrated to be due to their oxygen-loss, while the insulating nature of grain boundaries is due to their oxygen-stoichiometry.

## II. EXPERIMENT PROCEDURE

The CCTO thin films of 480 nm in thickness were grown on Pt/Ti/SiO$_2$/Si substrates by pulsed laser deposition (PLD) using a stoichiometry ceramic target. As reported elsewhere,[12] the CCTO thin films show the polycrystalline nature and have preferential orientation in (220). The film surface is crack-free and smooth, and is composed of uniformly distributed rectangular shaped grains. To measure the dielectric properties, the Pt top electrode with diameter of 0.25 mm were deposited by sputtering through a shadow mask, forming parallel-plate capacitors with bottom electrode. After deposition of the top electrodes, the films were given different post-annealing treatments in a tube furnace as described in Table I. The dielectric properties and IS measurements were carried out at room temperature after each post-annealing using a HP4294A impedance analyzer at 100 mV over the frequency range of 40 Hz to 10 MHz.

**III. RESULTS AND DISSCUSIONS**

Figure 1 shows the variations of the dielectric constant and dielectric loss as a function of frequency for films annealed in $N_2$ and $O_2$ atmospheres at different temperature for 8 h. The as-deposited capacitors show good dielectric properties, where the dielectric constant and dielectric loss values at 100 kHz were approximately 2300 and 0.085, respectively, which are comparable with those obtained in the epitaxial CCTO films grown on oxides substrate.[11] The dielectric behavior of the Pt/CCTO/Pt capacitors was very sensitive to the post-annealing atmosphere and temperature. Films annealed at 200 °C in nitrogen and oxygen showed no significant changes in the dielectric properties. However, capacitors with nitrogen post-annealed at 400 °C presented strong dielectric relaxation at low frequencies ($10^2$ Hz-$10^4$ Hz), while the subsequent oxygen post-annealing diminished the low frequency dielectric relaxation effect. When the capacitors were again post-annealed in nitrogen at 550 °C, the phenomenon of a larger dielectric relaxation was once more observed, while the oxygen atmosphere annealing reduced the relaxation but caused a steady fall in the dielectric constant.

Fig. 2 shows typical impedance spectra for the as-deposited and annealed CCTO thin films corresponding to Fig.1. The as-deposited CCTO thin films contain a large arc (hollow curves) with nonzero intercept at high frequencies in impedance complex plane plots, which suggests that the CCTO thin films is electrical inhomogeneous. The impedance data could be modeled using an equivalent circuit based on two parallel (resistor R, and capacitor C) elements connected in series. Similar to the bulk ceramic case,[4,17] the nonzero intercept indicates the presence of an arc for frequency higher than the maximum frequency measured ($10^7$ Hz), which is assigned as the contribution from grain interior, while the large semicircular arc at low frequencies is originated from the grain boundaries. The resistances of the grains and grain boundaries can be estimated from the intersects of the curves with the real part of the impedance Z' axis as 20 Ω [inset of Fig. 2 (a)] and 44 kΩ [ Fig. 2 (a)], which corresponds to the resistivities of 204 Ω cm and 450 kΩ cm, respectively. It can be noted that the values of resistivity are very close to those of the CCTO ceramic.[4] Thus, the room temperature IS data indicate that the polycrystalline CCTO thin films consist of insulating grain boundaries with semiconducting grains, and therefore, the high dielectric constant can be ascribed to the BLC mechanism.

The $Z^*$- plots for the CCTO thin films annealed at different stages show that the resistances of the grains and grain boundaries are very sensitive to the annealing atmosphere and temperature. When the films were annealed in nitrogen atmosphere at 200 °C, the resistance of the grains remained constant, while the resistance of the grain boundaries decreased from 44 kΩ to 14 kΩ. When the films were subsequently annealed in oxygen atmosphere at 200 °C, no obvious change was observed in the resistance of the grains, but the resistance of the grain boundaries increased. Similar behavior was also found in higher annealing temperatures, the resistance of the grain boundaries decreased in nitrogen atmosphere, will regain the resistence when the films were again post-annealing in oxygen atmosphere. Moreover, when the films were annealed in oxygen atmosphere at 400 °C, the resistance of the grains increased to 125 Ω and could be restored as before when the films subsequently annealed at 550 °C in nitrogen. Most important, when the films were annealed at 550 °C in oxygen, we found that the impendence data for the CCTO thin films only took the form of a single arc with the resistance value of 660 kΩ. At high frequencies (close to 10 MHz) the arc intersects with the Z" axis at a non-zero value. We assume that this arc was originated from the grain components, which could be checked by the fitting results discussed below. In addition, several points of the arc at low frequencies was observed, this was attributed to the grain boundaries.

Obviously, the resistances of the grains and grain boundaries of the CCTO films were tuned by changing annealing atmosphere and temperature. Under oxygen-absent (nitrogen) annealing atmosphere, the resistance of the grain boundaries decreases even at an annealing temperature of 200 °C, and this decrease of the resistance was enhanced when the annealing temperature increases. While the resistance of the grains keeps almost the same at about 25 Ω, irrelevant to the annealing temperature, as shown in the insert figures of Fig.2(b), 2(d) and 2(e), respectively. However, under oxygen annealing atmosphere, the reverse happened. First, the resistance of the grain boundaries was restored and even increased with the increase of annealing temperature. Second, the resistance of the grains also increased significantly after post-annealed in oxygen atmosphere.

The CCTO films crystalize at 720 °C, which is well above the present annealing temperatures. From XRD and SEM reasults, we found no structural changes for the samples experienced all the above post-annealing processes. Thus, the resistance variations was not from crystalline structural change. On the other hand, it is a well known fact that when the peroviskite oxide are subject to reduced oxygen atmosphere annealing, such as in nitrogen gas, oxygen vacancies are produced by the following process in some oxygen-stoichimetric region:

$$\mathbf{O_o}\text{(Oxygen ion at its normal site)} \longrightarrow \mathbf{V_o^{++}}\text{(Oxygen vacancy)} + \mathbf{2e^-} + \mathbf{1/2O_2} \quad (1)$$

This process can be reversed only after the samples with oxygen vacancies are annealed in oxygen atmosphere at proper temperatures. In the current study, when the film was annealed in nitrogen atmosphere, resistance decreases were only found in the grain boundaries, indicating that oxygen vacacies only formed in the grain boundaries but not inside the grains. This point demonstrated that the grain boundaries are originally close to oxygen-stoichimetric.

When the film annealed in oxygen atmosphere, the resistance of the grain boundaries restored or even increased. In addition, the resistance of grains also increased. This informs us that, when the films are post-annealed in oxygen atmosphere, the reverse reaction of Eq. (1) happened, and the oxygen vacancies in both grain boundaries and grains are compensated. Thus, in the grains, which were already oxygen deficient with semiconducting nature, the reaction of Eq. (1) is difficult to happen in reducing atmosphere. However, the reverse process of Eq.(1) is easy to happen when annealed in oxygen atmosphere. This leads to their resistance inert to the post-annealing in reducing atmosphere, and sensitive to the post-annealing in oxygen atmosphere. Here we note that in ordinary perovskite oxide polycrystalline films, such as BST films, the grains and grain boundaries are originally both oxygen-stoichimetric and insulating. When post-annealed in nitrogen atmosphere at proper temperatures, oxygen vacancies will be generated in both, resulting in the increase of the dielectric constant in the low-frequency region. Subsequent annealing in oxygen atmosphere will compensate these oxygen vacancies, causing the disappearance of the low-frequency dielectric dispersion[22,23] Thus, the decrease of the dielectric constant to a value lower than the original one can not occur. However, for the present CCTO polycrystalline films, the grains are originally deficient. In the last step of post-annealing in Table 1, not only the oxygen vacancies in the grain boundaries but also those in the grains were compensated, leading to the increase of the resistances of the grains and grain boundaries. Thus, the original semiconducting grains become partially insulating, resulting the increase of the effective thickness of the grain boundaries. Hence, the dielectric constant can be decreased even below the original value.

To analyze the details of the impedance response of the CCTO thin films, we may assign a model equivalent circuit. Conventionally, the simplest equivalent circuit is a series network of three parallel *RC* elements, which has been applied successfully in the CCTO ceramics,[4,17] as shown in Fig. 3(a). Each *RC* element represents one of the responses of grain, grain boundary and electrode interface, and will give rise to a semicircular arc in the impedance $Z^*$ plot. The grain and grain boundary responses can be identified by the magnitude of *C* associated with each arc, which are typically of the order of pF and nF, respectively.[21] It is not necessary to include an impedance element representing the electrodes when the contact resistance is small, as in the present experiment. For such a circuit, the complex impedance of each arc is given by:

$$Z^* = \frac{R}{(1+j\omega RC)} = \frac{R}{(1+j\omega \tau)} \qquad (2)$$

where $\tau = RC$, is the relaxation time constant of the *RC* element and $\omega$ is the angular frequency. In the complex impedance plane plot, the frequency at which the semicircular arc maximum occurs is determined by the relaxation time constant $\tau$, and can be described by:

$$\omega_{max} = \tau^{-1} = (RC)^{-1} \tag{3}$$

Using Eqs. (2) and (3), the elements of the equivalent *RC* circuit can be extracted form $Z^*$ plots, and the fitting results of the CCTO thin films was summarized in Table II. The curve in Fig. 2 is drawn according to the fitting results.

Because we used ideal resistors and capacitors in the *RC* equivalent circuit, each semicircle in the $Z^*$ plots should be perfect with its center on the real axis of the complex plane, just like the behavior of the as-deposited CCTO thin films. However, as shown in Fig. 2 (d-g), the experiment arc at high annealing temperatures could not yield a perfect semicircle, but a depressed one with its center below the real, which indicates that the electrical response of grain boundaries or grains in the annealed CCTO thin films exhibit a kind of distributing impedance response. The distributing factor in the equivalent circuit should be taken into account. In order to obtain better fitting, constant phase element (CPE) is introduced to replace the capacitor in the *RC* circuit as shown in Fig. 3(b). It is known that the impedance of CPE can be described as:[20]

$$Z_{CPE} = \left(\frac{1}{A}\right) \times (j\omega)^{-n} \tag{4}$$

Where $\omega$ is the angular frequency, $A$ and $n$ are constants and $0 \leq n \leq 1$. The CPE describes an ideal capacitor with $C = A$ for $n = 1$ and an ideal resistor with $R = 1/A$ for $n = 0$. Thus, Eq. (2) can be modified as:

$$Z^* = \frac{R}{\left(1+(j\omega)^n RA\right)} = \frac{R}{\left(1+(j\omega\tau)^n\right)} \tag{5}$$

where $\tau^n = RA$, and Eq. (3) can be expressed as:

$$\omega_{max} = \tau^{-1} = (RA)^{-1/n} \tag{6}$$

Using the above equations, the equivalent circuit elements can be fitted from the $Z^*$ plots. The triangle dots in Fig. 2 are the fitting result by using CPE model, and the fitting parameters are listed in Table II. It can be seen that the fitted impedance curves are in better agreement with experiment data than those using the *RC* model. Such phenomenon was also found in another BLC material: Li and Ti doped NiO.[24,25]

The fitting results also support our explanation about the variation of the dielectric properties. The distributed electrical response is always associated with the distribution of relaxation times, and this distribution is a result of the distributed elements caused by the presence of electrical inhomogeneities in the samples.[20] In the present work, the as-deposited CCTO thin films consists two types of electrical components: grains and the grain boundaries. From Table II and Fig. 2, we can find that both *RC* model and CPE model can fit the IS data very well for the as-deposited CCTO thin films. The CPE model with a value for n = 0.99 is actually the *RC* model, which indicates that the components of grains and grain boundaries in the as-deposited CCTO thin films are nearly electrical homogeneous. Similar results were also found in the CCTO thin films annealed at 200 $^o$C in oxygen or nitrogen atmosphere. However, at higher annealing temperatures, distributed impedance response in grain boundaries or grains becomes evident. Obviously, when the film was post-annealed at higher temperature, the reaction (1) and its reverse process is easy to happen. In addition, we propose that, in some regions of the film, such as the interface of the grains and grain boundaries, the rate of the reaction (1) is different from the others, which leads to the electrical properties of these regions being different one from the other. This can generate electrical inhomogeneities in the components of the sample. So the *RC* model could not describe the experiment data while the CPE model is more suitable for the post-annealed polycrystalline CCTO thin films. Moreover, we also found that when the films are annealed in oxygen atmosphere at 550 $^o$C, the capacitance of the observed semicircle is in the order of pF, which is the value close to that of the grains, confirming our assumption that the large arc in Fig. 2(g) was originated from the grain components.

## V. SUMMARY AND CONCLUSIONS

In summary, Polycrystalline CCTO thin films were fabricated on Pt/Ti/SiO$_2$/Si(100) substrates by PLD. Using IS technique, the electrical microstructures of the as-deposited and post-annealed polycrystalline CCTO thin films were studied. Our study demonstrated that the as-deposited polycrystalline

CCTO film consists of insulating grain boundaries with semiconducting grains, indicating that the high-dielectric-constant is attributed to the BLC effects. A traditional RC model does not give a good fit to the experimental data at high annealing temperatures. A modified model, which the capacitor in the *RC* circuit replaced by a constant phase element, was then used to fit the IS data and excellent agreement between the calculated and measured curves was obtained. Moreover, it was found that when the films are annealed in different post-annealing atmosphere and temperature, the electric properties of the grains and the grain boundaries are different, which means that the dielectric properties of the CCTO thin films can be tuned through different annealing processes. Our results also provided additional evidence for the nature of the semiconductivity grains and insulating grain boundaries.

## VI. ACKNOWLEDGEMENTS

The authors acknowledge the financial support of Natural Science Foundation of China (Grant No. 10204016)

**References:**

[1] C. C. Homes, T. Vogt, S. M. Shapiro, S. Wakimoto, and A. P. Ramirez, Science 2**93,** 673 (2001).

[2] M. A. Subramanian, L. Dong, N. Duan, B. A. Reisner, and A. W. Sleight, J. Solid State Chem. **151,** 323 (2000).

[3] A. P. Ramirez, M. A. Subramanian, M. Gardel, G. Blumberg, D. Li, T.Vogt, and S. M. Shapiro, Solid State Commun. **115,** 217 (2000).

[4] T. B. Adams, D. C. Sinclair and A. R. West, Adv. Mater. **14**, 1321 (2002).

[5] A. Kotizsch, G. Blumberg, A. Gozar, B.Dennis, A. P. Ramirez, S. Trebst, and S. Wakimoto, Phys. Rev. B **65**, 052406 (2002).

[6] L. X. He, J. B. Neaton, M. H. Cohen, and D. Vanderbilt, Phys. Rev. B **65**, 214112 (2002).

[7] N. Kolev, R. P. Bontchev, A. J. Jacobson, V. N. Popov, V. G.Hadjiev, A. P. Litvinchuk, and M. N. Iliev, Phys. Rev. B **66**, 132102(2002).

[8] P. Lunkenheimer, V. Bobnar, A. V. Pronin, A. I. Ritus, A. A. Volkov, and A. Loidl, Phys. Rev. B **66,** 052105 (2002).

[9] C. C. Homes, T. Vogt, S. M. Shapiro, S. Wakimoto, M. A. Subramanian, and A. P. Ramirez, Phys. Rev. B **67,** 092106 (2003).

[10] Y. Lin, Y. B. Chen, T. Garret, S. W. Liu, C. L. Chen, L. Chen, R. P. Bontchev, A. Jacobson, J. C. Jiang, E. I. Meletis, J. Horwitz, and H. D. Wu, Appl. Phys. Lett. **81**, 631 (2002).

[11] W. Si, E. M. Cruz, P. D. Johnson, P. W. Barnes, P. Woodward, and A. P. Ramirez, Appl. Phys. Lett. **81**, 2058 (2002).

[12] L. Fang and M. R. Shen, Thin Solid Films 440, 60 (2003).

[13] R. L. Nigro, R. G. Toro, G. Malandrino, M.Bettinelli, A, Speghini and I. L. Fragala, Adv. Mater. **16**, 891 (2004).

[14] A. Tselev, C. M. Brooks, S. M. Anlage, H. Zheng, L. S. Riba, R. Ramesh and M. A. Subramanian, (2003) cond-mat/0308057

[15] M. H. Cohen, J. B. Neaton, L. X. He, and D. Vanderbilt, J. Appl. Phys. **94,** 3299 (2003).

[16] A. P. Ramirez, G. Lawes, V. Butko, M. A. Subramanian, and C. M. Varma, (2002) cond-mat/0209498.

[17] D. C. Sinclair, T. B. Adams, F. D. Morrison, and A. R. West, Appl. Phys. Lett. **80**, 2153 (2002).

[18] C. F. Yang, Jpn. J. Phys., Part 1 **36,** 188 (1997).

[19] L. Fang, M. R. Shen and W. W. Cao, J. Appl. Phys. **95,** 6483 (2004).

[20] J. Ross Macdonald, *Impedance Spectroscopy* (Wiley, New York, 1987).

[21] D. C. Sinclair and A. R. West, J. Appl. Phys. **66**, 3850 (1989).

[23] F. M. Pontes, E. R. Leite, E. Longo, J. A. Varela, E. B. Araujo and J. A. Eiras, Appl. Phys. Lett. **76**, 2433 (2000).

[23] M. R. Shen, Z. G. Dong, Z. Q. Gan, S. B. Ge and W. W. Cao, Appl. Phys. Lett. **80**, 2538 (2002).

[24] J. B. Wu, C. W. Nan, Y. H. Lin, and Y. Deng, Phys. Rev. Lett. **89**, 217601 (2002).

[25] J. B. Wu, J. Nan, C. W. Nan, Y. Lin, Y. Deng and S. Zhao, Mater. Sci. Eng., B **99**, 294 (2003).

**Table captions**

Table I. Post-annealing conditions of the each step.

Table II. The fitting results by *RC* model and CPE model.

**Figure Captions:**

Figure 1. Frequency dependence of the dielectric constant $\varepsilon$ (a) and the loss $tan\delta$ (b) for the as-deposited CCTO thin films and the annealed thin films in different step.

Figure 2. Impedance complex plane plot for the CCTO thin films corresponding to figure 1. The insets show an expanded view of the high frequency data close to the origin. Hollow circles represent experiment data, curves represent the fitting results using the *RC* model and triangles represent the fitting results using the CPE model, respectively. Filled symbols indicate selected frequencies.

Figure 3. (a) The *RC* equivalent circuit model. $R_g$, $C_g$; $R_{gb}$, $C_{gb}$; and $R_e$, $C_e$ are resistance and capacitances associated with grains, grains boundaries and the electrodes, respectively. (b) The CPE equivalent circuit model. $CPE_g$, $CPE_{gb}$, and $CPE_e$ represent the CPE element of grains, grains boundaries and the electrodes, respectively.

Table I. Liang Fang

| Step | 1 | 2 | 3 | 4 | 5 | 6 |
|---|---|---|---|---|---|---|
| Post-annealing Conditions | 200 °C $N_2$, 8 h | 200 °C $O_2$, 8 h | 400 °C $N_2$, 8 h | 400 °C $O_2$, 8 h | 550 °C $N_2$, 8 h | 550 °C $O_2$, 8 h |

Table II. Liang Fang

| | | $R_{gb}$ (Ω) | $C_{gb}$ (F) | $R_g$ (Ω) | $C_g$ (F) |
|---|---|---|---|---|---|
| RC Model | as-deposited | 44.0×10³ | 2.58×10⁻⁹ | 20.0 | — |
| | N₂ 200°C | 14.0×10³ | 2.52×10⁻⁹ | 25.0 | — |
| | O₂ 200°C | 80.0×10³ | 2.65×10⁻⁹ | 20.0 | — |
| | N₂ 400°C | 7.6×10³ | 2.01×10⁻⁹ | 27.0 | — |
| | O₂ 400°C | 90.0×10³ | 1.18×10⁻⁹ | 125.0 | — |
| | N₂ 550°C | 0.6×10³ | 2.84×10⁻⁹ | 27.0 | — |
| | O₂ 550°C | — | — | 6.6×10⁵ | 9.70×10⁻¹² |

| | | $R_{gb}$ (Ω) | $n_{gb}$ | $A_{gb}$ (Ω⁻¹) | $R_g$ (Ω) | $n_g$ | $A_g$ (Ω⁻¹) |
|---|---|---|---|---|---|---|---|
| CPE Model | as-deposited | 44.0×10³ | 0.999 | 2.49×10⁻⁹ | 20.0 | — | — |
| | N₂ 200°C | 14.0×10³ | 0.999 | 2.59×10⁻⁹ | 25.0 | — | — |
| | O₂ 200°C | 81.0×10³ | 0.995 | 2.58×10⁻⁹ | 20.0 | — | — |
| | N₂ 400°C | 8.0×10³ | 0.920 | 4.25×10⁻⁹ | 27.0 | — | — |
| | O₂ 400°C | 104.0×10³ | 0.903 | 2.13×10⁻⁹ | 125.0 | — | — |
| | N₂ 550°C | 0.7×10³ | 0.881 | 9.84×10⁻⁸ | 27.0 | — | — |
| | O₂ 550°C | — | — | — | 6.7×10⁵ | 0.950 | 1.74×10⁻¹¹ |

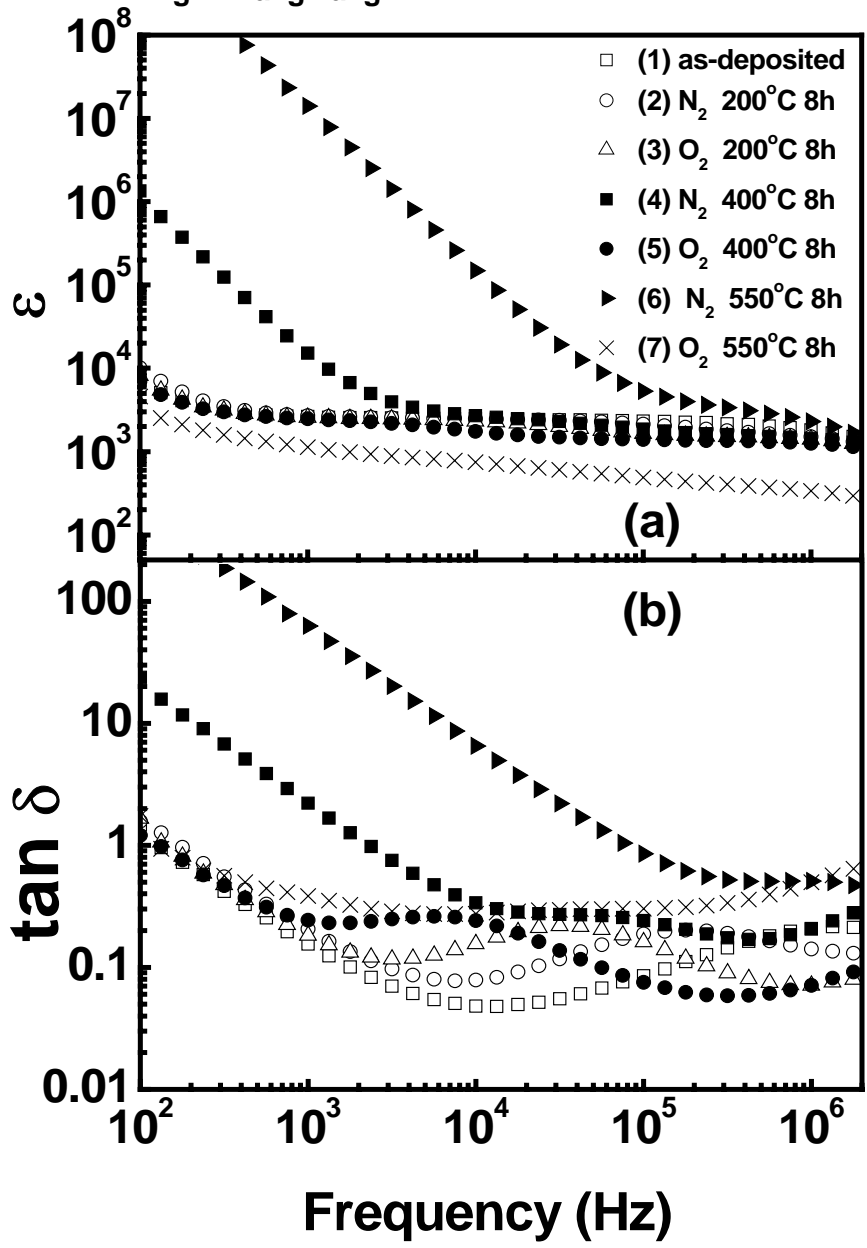

Fig. 1 Liang Fang

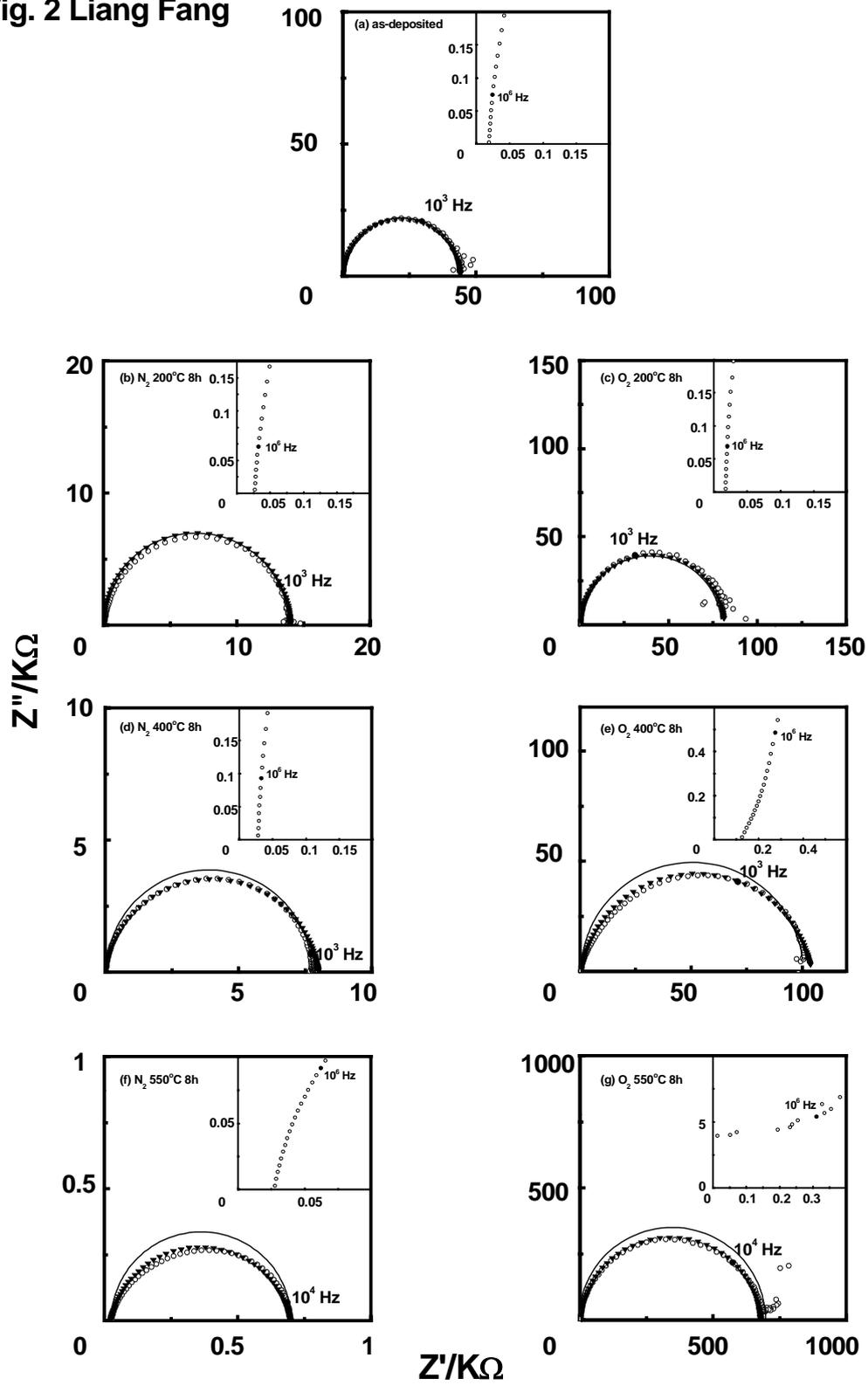

Fig. 2 Liang Fang

**Fig. 3 Liang Fang**

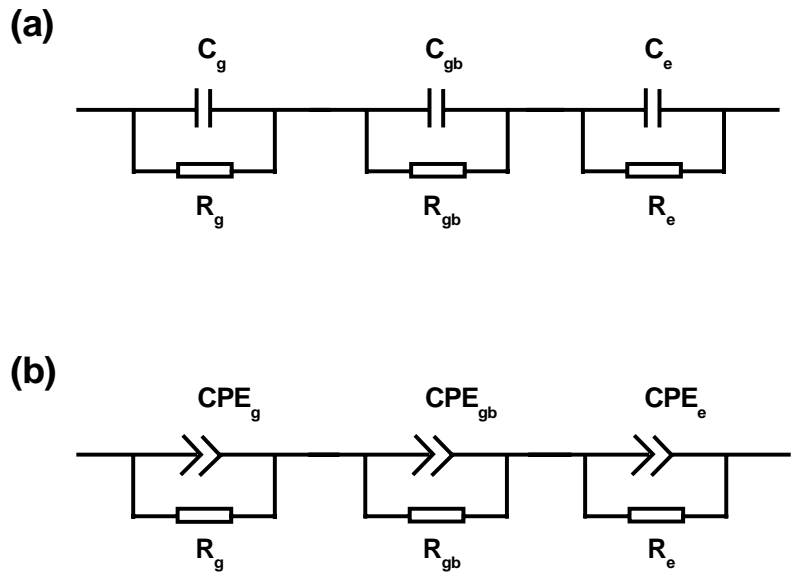